\documentclass[a4paper]{jpconf}
\usepackage{graphicx}
\usepackage{xspace}
\usepackage{booktabs,multirow,dcolumn,bigdelim}

\newcommand {\auau}{\mbox{Au$+$Au}\xspace}

\newcommand {\pp}{\mbox{$p$$+$$p$}\xspace}

\newcommand{\dAu}{\mbox{$d$$+$Au}\xspace}

\begin{document}
\title{The Physics of sPHENIX}

\author{Anne M. Sickles, for the PHENIX Collaboration}

\address{Physics Department, Brookhaven National Laboratory, Upton, NY 11973}

\ead{anne@bnl.gov}

\begin{abstract}
Jet related observables have been some of the most powerful and exciting probes for understanding
the matter produced in ultra-relativistic heavy ion collisions.  Full jet reconstruction
was begun at RHIC, and the LHC experiments have shown the power and kinematic reach of these
observables.  Here we discuss the sPHENIX detector and physics program which aims to bring
full calorimetric based jet reconstruction to RHIC in order to explore the temperature
dependence of the strongly interacting Quark Gluon Plasma.
\end{abstract}

\section{Introduction}

In 2010 the RHIC collaborations were charged to set out their plan sfor
the next decade of RHIC running.   After much discussion the PHENIX Collaboration
decided that the most exciting heavy ion physics over the coming years  was hard probe
physics (jets, quarkonia and heavy flavor), but that quality measurements
of these observables over the next decade were incompatiable with the existing PHENIX central detectors
which have small acceptance and lack hadronic calorimetry.  What emerged from 
these considerations was a more radical plan
of replacing the current central arms of PHENIX with a compact calorimeter and solenoid.
This idea generated a lot of interest and the PHENIX Collaboration has recently written
a proposal laying out the physics case and a planned design~\cite{Aidala:2012nz}.

\section{Physics of sPHENIX}

The goal of the sPHENIX Upgrade to the PHENIX experiment is to make calorimetric
jet measurements at RHIC in order to study the Quark Gluon Plasma in the temperature
region near the critical temperature.

\begin{figure}
\includegraphics[width=0.45\textwidth]{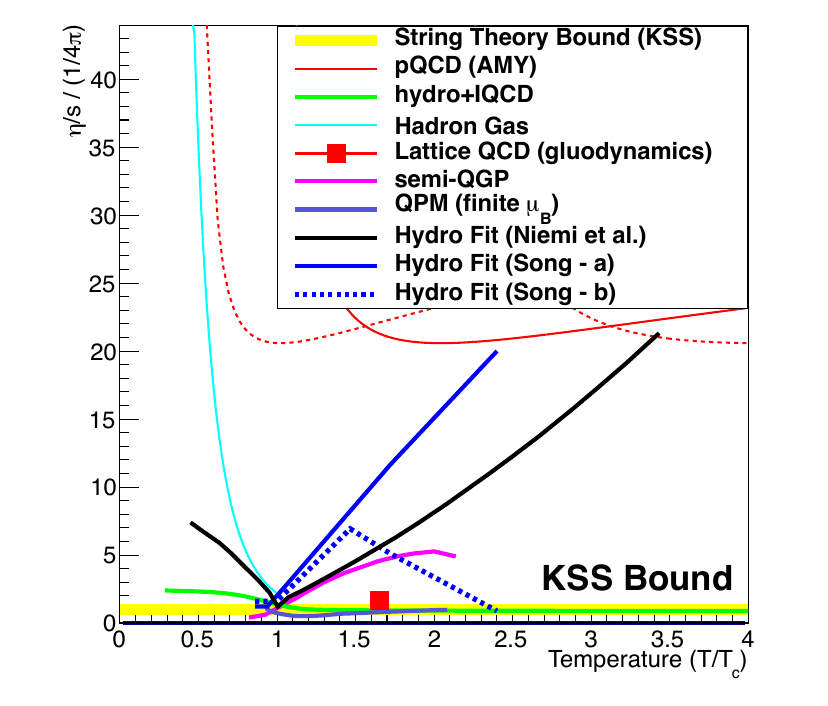}
\includegraphics[width=0.50\textwidth]{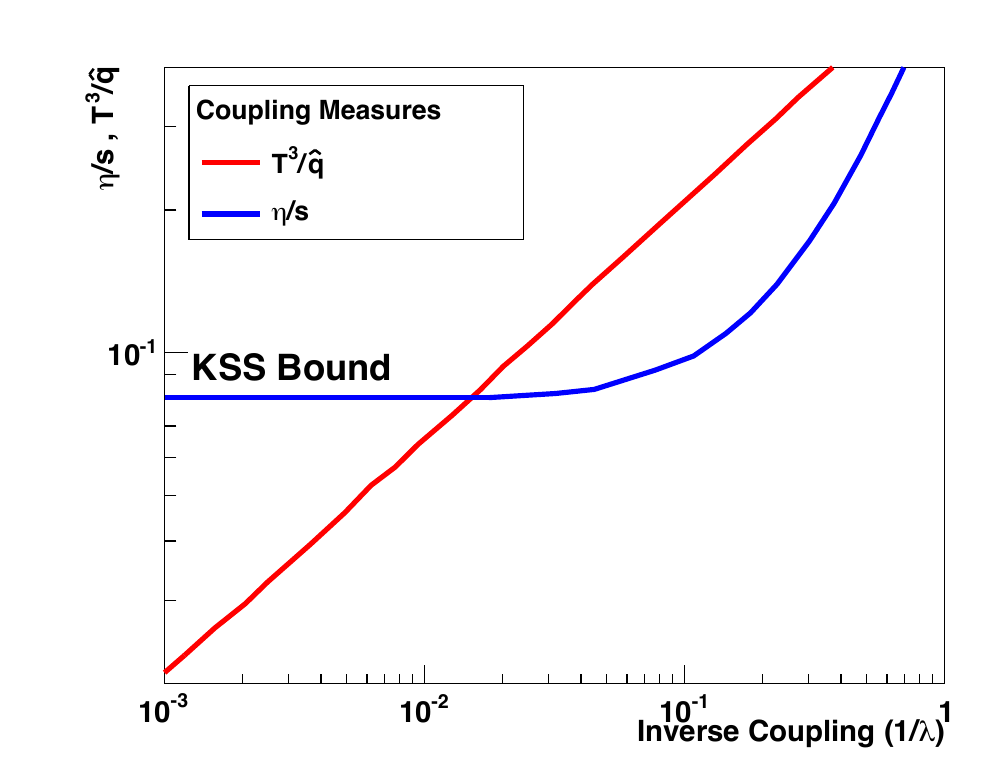}
\label{fig:tdep}
\caption{(left)$\eta/s$ as a function of temperature.  The red curves show the perturbative result from 
AMY~\cite{Arnold:2003zc}.  The blue and black curves show parameterizations of $\eta/s$ which
have been used in hydrodynamical models to reproduce existing RHIC and LHC 
results~\cite{Song:2011qa,Niemi:2011ix}. 
(right) $\frac{T^3}{\hat{q}}$ (red) and $\eta/s$ (blue) as a function of the inverse 't Hooft
coupling, $\lambda$~\cite{Majumder:2007zh}.}
\end{figure}

The discovery of the extremely low shear viscosity to entropy density ratio,
$\eta/s$, 
established that RHIC created the QGP in a regime that was 
characterized by strong coupling rather than a weakly coupled gas of 
quarks and gluons.  The $\eta/s$ values required to reproduce the experimental
flow measurements are small~\cite{Luzum:2012wu} and within a factor of a few of the conjectured
lower quantum bound~\cite{Kovtun:2004de}.  The left panel of Figure~\ref{fig:tdep} shows the
lower quantum bound and the perturbative calculation for $\eta/s (T)$.  
It is not known how $\eta/s$ evolves from the near minimum value near $T_c$ to the
perturbative value at very high temperatures.  Flow data are not able to constrain
the temperature dependence adequately.

One interesting theoretical development is the identification of a connection
between $\eta/s$ and $\hat{q}$, the transverse momentum broadening per unit length
of a fast parton as it traverses the QGP~\cite{Majumder:2007zh}.  At
weak coupling:
\begin{equation}
\frac{\eta}{s} = 1.25\frac{T^3}{\hat{q}}
\end{equation}
while at strong coupling:
\begin{equation}
\frac{\eta}{s} \gg \frac{T^3}{\hat{q}}.
\end{equation}
The authors of Ref.~\cite{Majumder:2007zh} state that $T^3/\hat{q}$ ``is a more
broadly valid measure of the coupling strength of the medium than $\eta/s$."
This is illustrated in the right panel of Figure~\ref{fig:tdep} which shows
both $T^3/\hat{q}$ and $\eta/s$ as a function of the inverse coupling.  As
the coupling becomes large $\eta/s$ saturates at the quantum bound, $1/4\pi$,
while $T^3/\hat{q}$ retains its sensitivity to further increases in the coupling strength.

The key to understanding the coupling in the QGP is to measure both $\eta/s$ and
$\hat{q}$, and their temperature dependences, independently.  
This requires high quality jet measurements at RHIC collision energies.  Each collision
evolves from its maximum initial temperature down as the system expands and cools.
Since $\hat{q}$ depends on $T^3$ in pQCD the quenching is dominated by the highest temperatures
in the collision.  Thus, the way to study the temperature dependence is the vary
the maximum initial temperature by varying the collision energy.  Measurements at the
LHC alone cannot do this.

\section{sPHENIX Design}

The jet performance desired for sPHENIX drives the design considerations.  Full calorimeter coverage,
both electromagnetic and hadronic is required.  The detector needs to have full azimuthal
coverage over $|\eta|<$1.  The currently existing Silicon Vertex Detector (VTX) is to remain for tracking,
with possible additional silicon layers added at larger radius to improve the momentum resolution
and improve the track finding ability.

\begin{figure}
\centering
\includegraphics[width=0.7\textwidth]{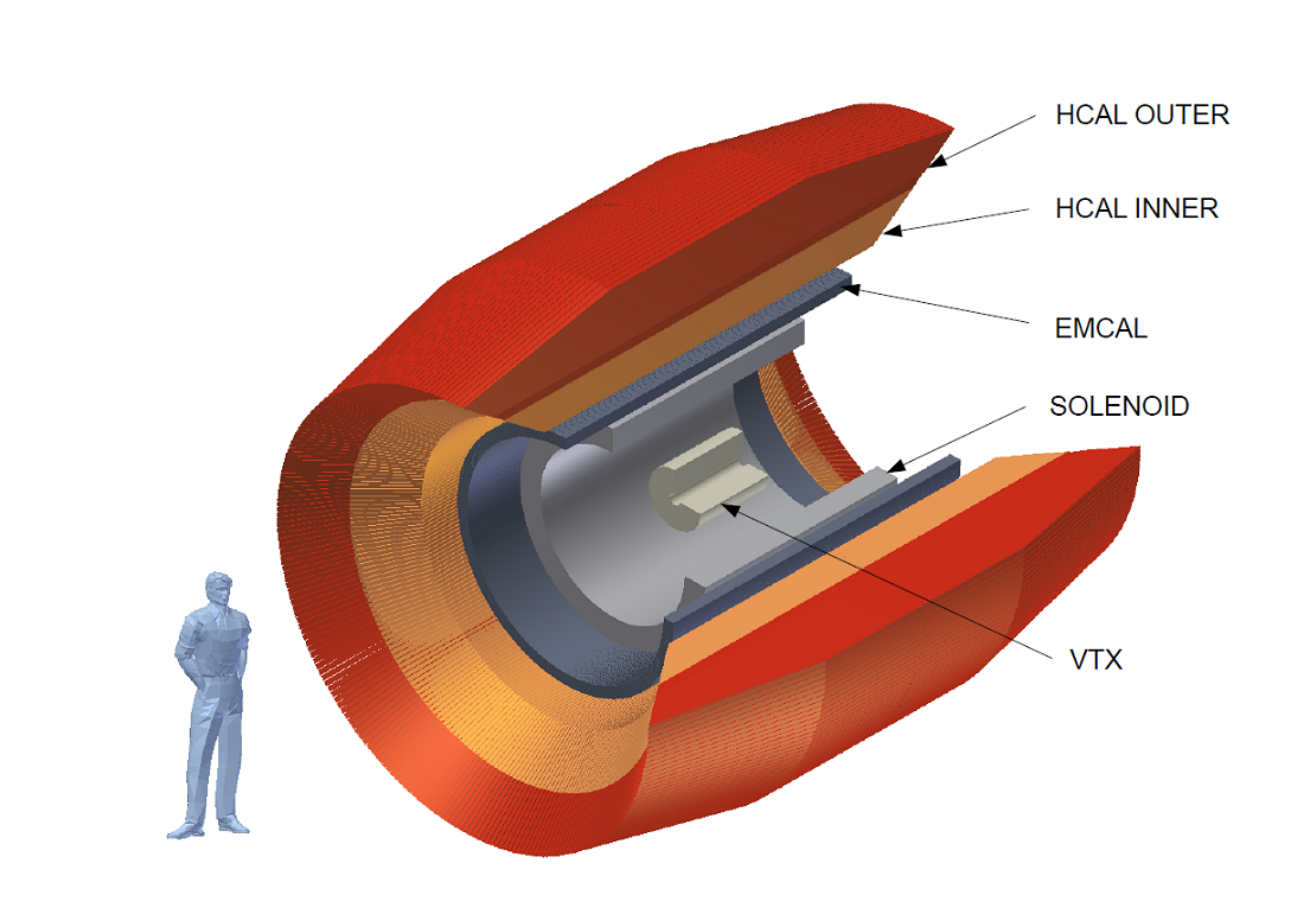}
\label{fig:sphenix}
\caption{A cut-way view of the sPHENIX detector.}
\end{figure}

A cut-away view of the detector is shown in Figure~\ref{fig:sphenix}.
The design is for a thin superconducting solenoid with a 2T field at a radius of 70 cm.  Behind this
is a silicon tungsten electromagnetic calorimeter followed by an iron scintillator hadronic calorimeter.
Using tungsten as the absorber allows the electromagnetic calorimeter to be very compact; the calorimeter
is about 10 cm thick.  Geant 4 single particle simulations give an energy resolution of 14.2\%$/\sqrt{E}$
+0.7\%.  

\section{Jet Reconstruction Performance at RHIC}
\begin{figure}
\centering
\includegraphics[width=0.45\textwidth]{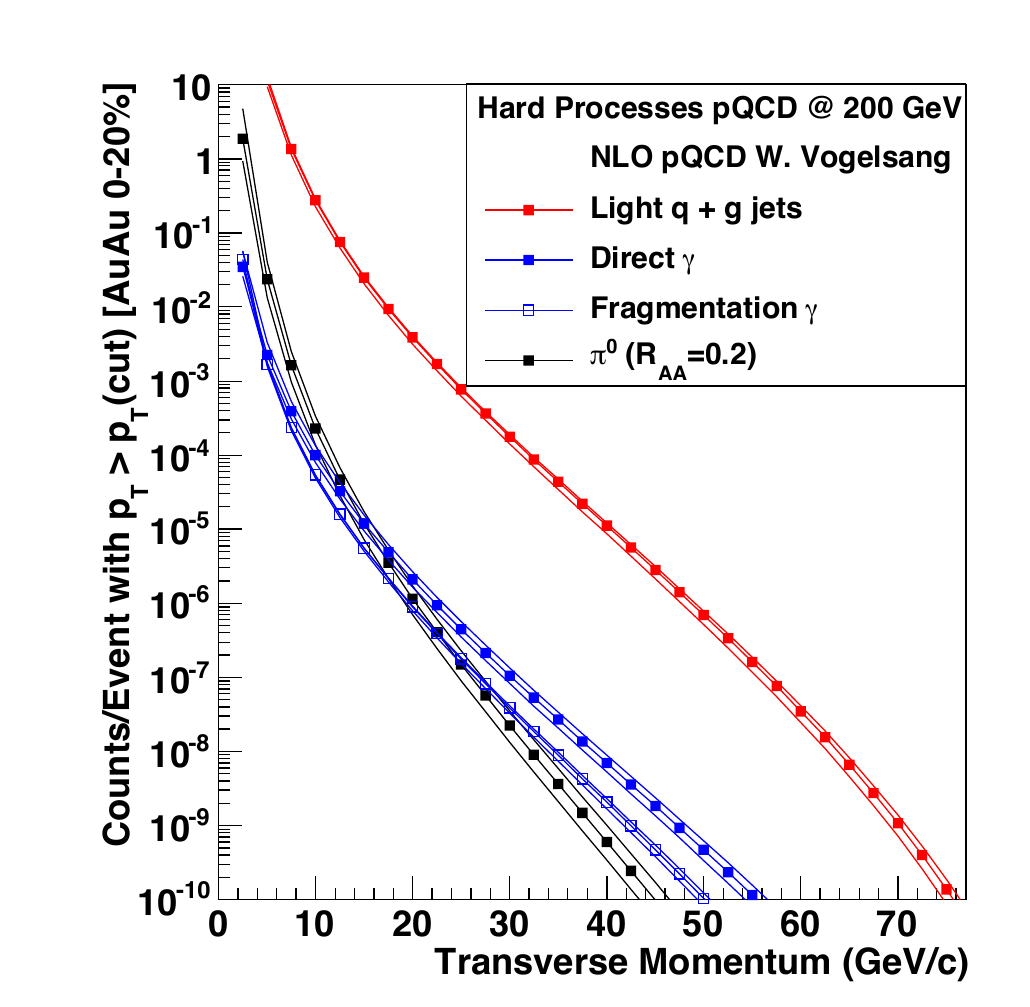}
\includegraphics[width=0.46\textwidth]{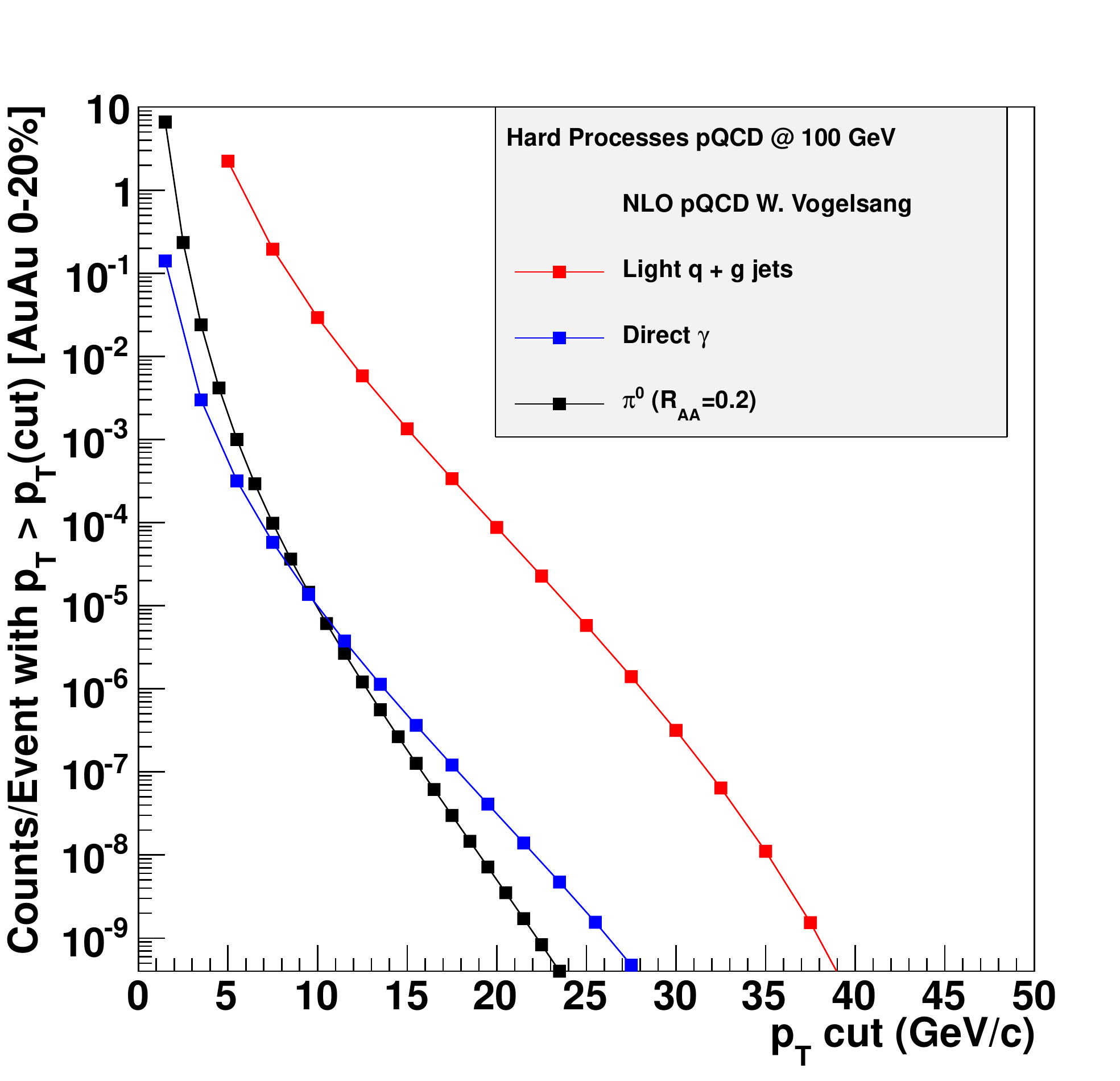}
\label{fig:rates}
\caption{NLO jet, $\pi^0$ and photon rates~\cite{Vogelsang:NLO} at $\sqrt{s_{NN}}$=200~GeV (left) and 
$\sqrt{s_{NN}}$=~100~GeV (right).  }
\end{figure}

The rates for jet production at RHIC given expected the machine luminosity and expected 20 week/year runs are such that 
sPHENIX would have access to huge numbers of jets up to approximately 60~GeV/c for
$\sqrt{s_{NN}}$=~200~GeV.  Figure~\ref{fig:rates} (left)
shows the pQCD rates for jet, photon and $\pi^0$ production in central Au+Au events.  sPHENIX would be able
to sample approximately 50B Au+Au events in a year.  The numbers of jets and
photons above various $p_T$ cuts are shown in Table~\ref{tab:nlo_jetrates}.  Over 80\%
of the single jets that sPHENIX 
also have the opposing jet  within the sPHENIX acceptance.  Jet rates
at $\sqrt{s_{NN}}$=~100~GeV~\cite{Vogelsang:NLO} are shown in Figure~\ref{fig:rates} (right)
and one 20 week run would yield $10^5$ jets with $p_T >$~20~GeV.

For $p_T>$~20~GeV the yield of direct photons surpasses that of photons from $\pi^0$ decay in
Au+Au collisions.  As can be seen in Table~\ref{tab:nlo_jetrates} there are abundant 
direct photons at RHIC in this region.  Because the photon does not interact via the
strong force, it decouples from the medium after it is created.  sPHENIX will be able
to make measurements of both photon-jet and photon-hadron correlations to probe
energy loss in this region.

\begin{table}[!hbt]
  \centering
  \begin{tabular}[c]{r l l l}
    & \multicolumn{1}{c}{{\renewcommand{\arraystretch}{1.2}
\begin{tabular}[t]{@{}c@{}}\auau\\(central 20\%)\end{tabular}} } &
\multicolumn{1}{c}{\pp} & \multicolumn{1}{c}{\dAu} \\
    \toprule
    \multirow{2}{*}{$ > 20$\,GeV} & $10^7$ jets & $10^6$ jets & $10^7$ jets \\
    & $10^4$ photons & $10^3$ photons & $10^4$ photons \\
    \midrule
    \multirow{2}{*}{$ > 30$\,GeV} & $10^6$ jets & $10^5$ jets & $10^6$ jets \\
    & $10^3$ photons & $10^2$ photons & $10^3$ photons \\
    \midrule
    $ > 40$\,GeV & $10^5$ jets & $10^4$ jets & $10^5$ jets \\
    \midrule
    $ > 50$\,GeV & $10^4$ jets & $10^3$ jets & $10^4$ jets 
  \end{tabular}
  \caption{Table of jet rates at $\sqrt{s_{NN}}$=~200~GeV for different systems. Each column shows
    the number of jets or direct photons that would be measured within
    $|\eta|<1$ in one 20 week running period.\label{tab:nlo_jetrates}}
\end{table}

\begin{figure}
\includegraphics[width=0.40\textwidth]{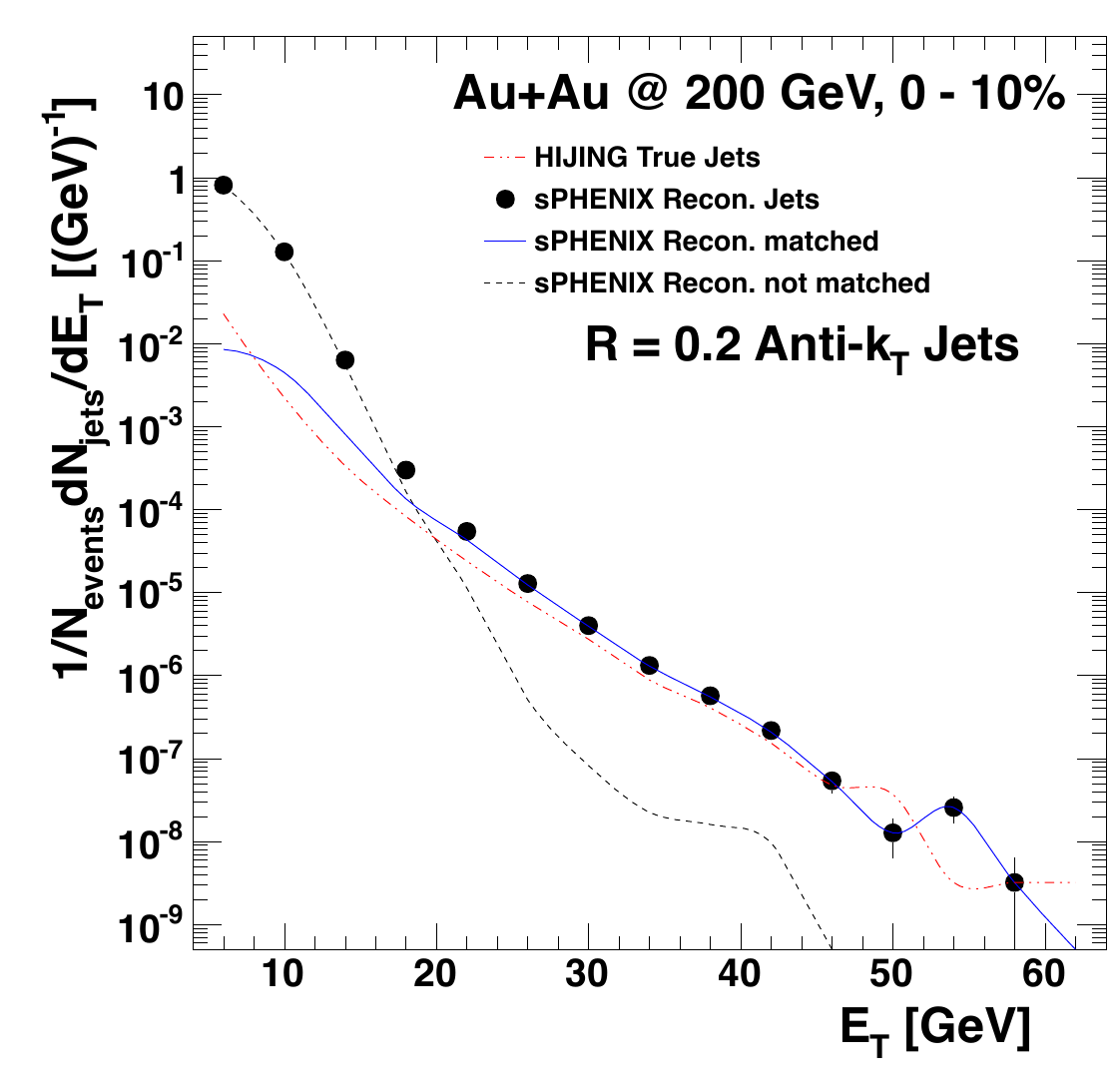}
\includegraphics[width=0.50\textwidth]{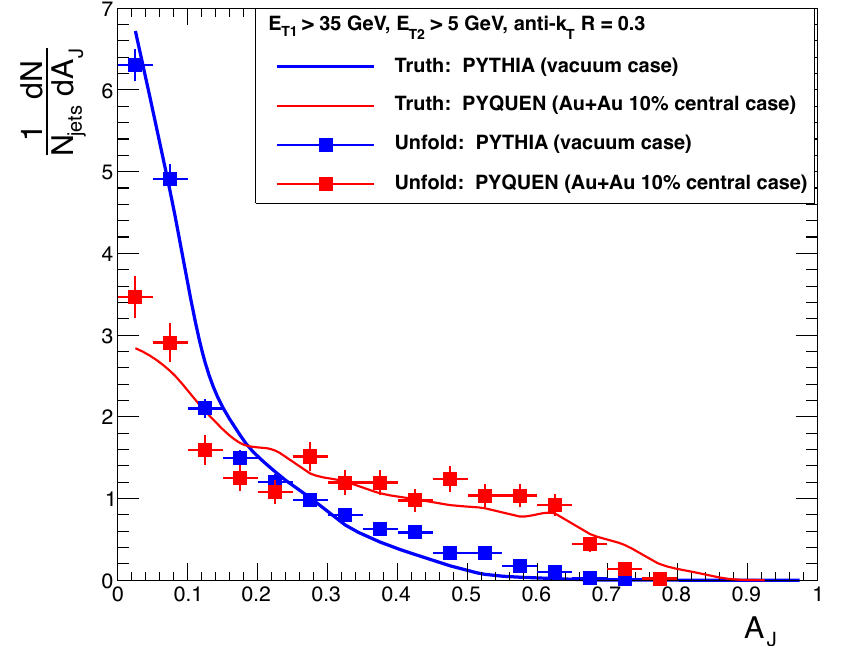}
\label{fig:jet_performance}
\caption{(left) Reconstructed single jet spectra for anti-$k_{T}$ R=0.2 jets in central 
HIJING~\cite{Gyulassy:1994ew} events.  The reconstructed jets are shown as points.  Those 
reconstructed jets matched with a true jet from the HIJING event are shown with the blue line
and jets not matched with a true jet (i.e. fake jets) are shown with the black dashed line.
For jet $E_{T} >$~20~GeV the matched jets dominate over the unmatched jets.  Figure is from 
Ref~\cite{Hanks:2012wv}. (right) Asymmetry, $A_{J}$, for PYTHIA~\cite{pythia} (blue)
and PYQUEN~\cite{Lokhtin:2005px} (red) for dijets embedded in 
central HIJING events.  The truth information is shown as lines and the reconstructed and
unfolded results as shown as points.  Figure is from Ref.~\cite{Aidala:2012nz}. }
\end{figure}

Numerous studies have been done to establish the feasibility of reconstructing jets
at $\sqrt{s_{NN}}$=~200~GeV in sPHENIX.  A large HIJING study was done in order to 
evaluate the separation of true jets from fake jets (background 
fluctuations)~\cite{Hanks:2012wv} in an ideal calorimeter.  Results for anti-$k_T$ R = 0.2 jets are
shown in Figure~\ref{fig:jet_performance} (left).  For jets with $E_T >$~20~GeV
true jets dominate over fake jets.  For larger jet radii the crossing 
point is at higher $E_T$, but still within the range that sPHENIX expects
to have statistics for.  

Dijet asymmetry measurements have been used extensively at the LHC.  In heavy ion collisions,
the large jet quenching decreases the fraction of symmetric (balanced) dijets and increases
the fraction of unbalanced dijets.  In order to estimate how well sPHENIX would be able
to distinguish these scenarios we embedded PYTHIA p+p events into central HIJING events and
reconstructed the jet asymmetry, $A_J$.  We also did the same with PYQUEN events, where jet
quenching is applied to PYTHIA events.  The results are shown in the right
panel of Figure~\ref{fig:jet_performance}.  The unfolded results for both the
PYTHIA and PYQUEN samples are in agreement with the initial truth asymmetry distributions.

\section{sPHENIX Upgrades}
As discussed above,
the sPHENIX proposal in Ref.~\cite{Aidala:2012nz} includes a solenoid and electromagnetic and
hadronic calorimetry.  This is appropriate for jet and direct photon measurements.  However,
other very interesting probes, such as separated upsilon states and heavy flavor jets will require additional
detectors.  There are plans for additional tracking layers beyond the existing VTX and a 
preshower detector that will be needed for electron identification.

The physics made available by these upgrades is extremely important
and the goal is to have these in place at the same time as the rest of sPHENIX.  Here we highlight one 
example, heavy flavor jets.  Heavy quarks, especially bottom, were expected to lose much less energy
than light quarks due to the dead cone effect~\cite{Dokshitzer:2001zm} suppressing gluon radiation.
However, results from both RHIC and the LHC have shown evidence for substantial energy
loss of both charm and bottom quarks~\cite{Adare:2006nq,Tlusty:2012ix,Nguyen:2012yx}.  

\begin{figure}
\centering
\includegraphics[width=0.7\textwidth]{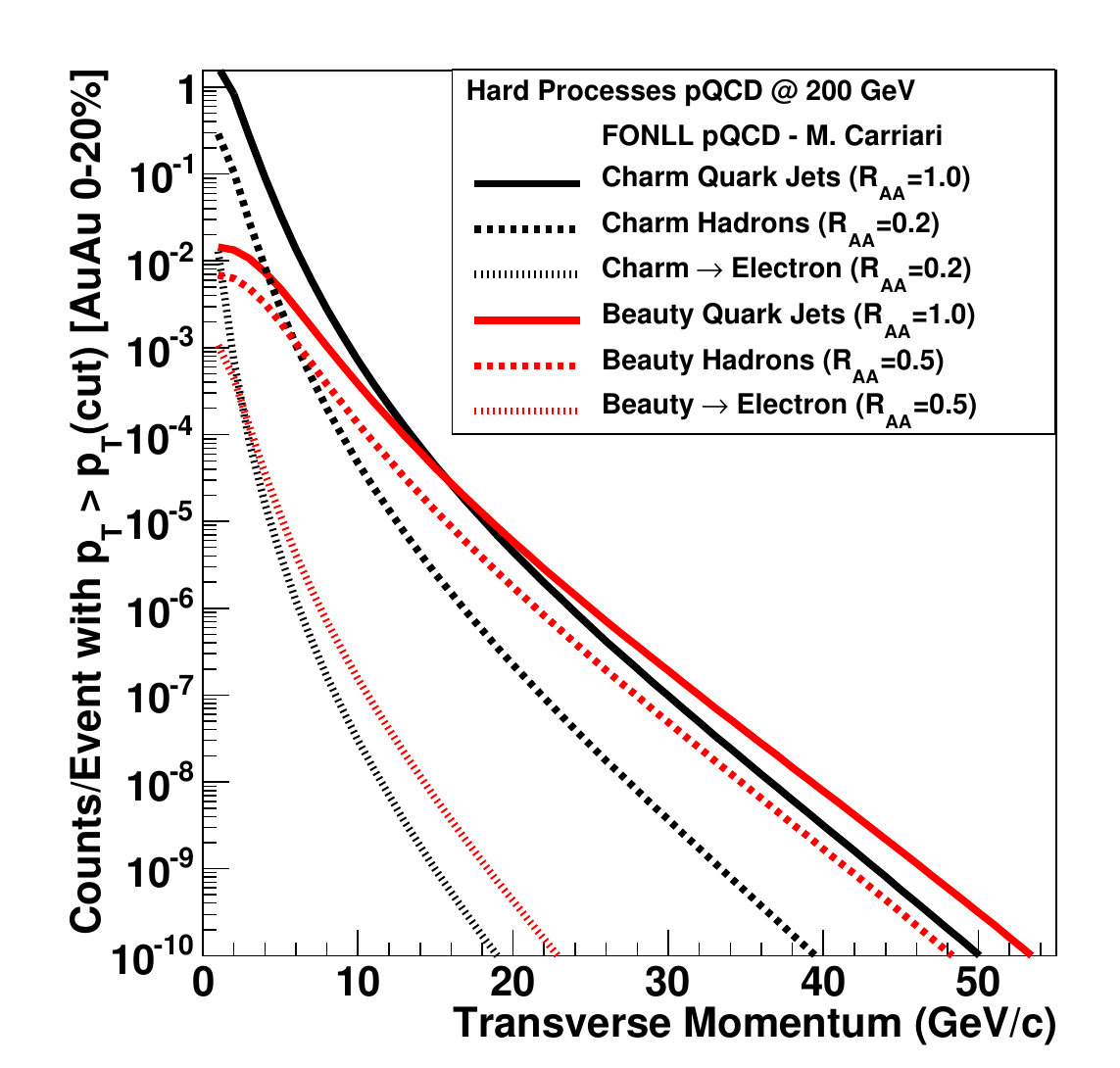}
\label{fig:hqrates}
\caption{Fixed order next-to-leading log (FONLL)~\cite{Cacciari:2005rk} results for 
heavy quark jet (solid lines), heavy hadron (thick dashed lines) and electron from semileptonic
heavy meson decay (thin dashed lines)~\cite{Cacciariprivate}.  Charm (black) and bottom (red) results are shown.
Hadron and electron rates have been reduced by the assumed $R_{AA}$ values shown in the figure.}
\end{figure}

If sPHENIX were to be capable of identifying heavy quark jets this would extend the
$p_T$ range of heavy quark measurements at RHIC significantly.  Figure~\ref{fig:hqrates}
shows that there are accessible rates for heavy quark production for $p_T>$~30~GeV/c.
The constraints from such measurements would also be greatly improved due to the ability
to constrain the kinematics from jet measurements which is not possible with electron or
heavy meson measurements.  Heavy quark jet measurements are a crucial part of understanding
hard physics and measurements are necessary both at RHIC and the LHC.

\section{Conclusions}

The sPHENIX detector will provide the first fully calorimetric jet measurements at RHIC.  These
measurements are crucial to understanding the behavior of fast partons in the QGP and
the properties of the plasma in a region where the coupling might be the strongest.

We have done simulations that show that anti-$k_T$ R = 0.2 jets can be cleanly measured for $E_T>$~20~GeV
with no additional fake jet rejection.  Applying fake jet rejection techniques, already being used at the
LHC~\cite{Aad:2012is} will decrease the jet energies and increase the jet sizes which are accessible. 

The sPHENIX design exploits technological advances in the years since PHENIX was constructed.  The 
sPHENIX design also offers the ability to add additional detectors in the central region which
will enable new key physics, such as heavy quark jets and separated upsilon states.  Plans
are also underway to instrument the forward region with emphasis on spin, asymmetric collisions and
future eRHIC running.

\section{References}

\bibliographystyle{iopart-num}
\bibliography{sickles_hq12}

\end{document}